\documentclass[12pt]{iopart}
\usepackage{setstack}
\usepackage{iopams}
\usepackage{mathrsfs}
\usepackage{amsthm}
\newcommand{\field}[2]{\mathbb{#1}^{#2}}
\newcommand{\cinf}{C^{\infty}_{0}}
\newcommand{\Hh}{\widehat{H}_{\hbar}}
\newcommand{\Heps}{\widehat{H}_{\varepsilon}}
\newcommand{\Hbo}{\widehat{H}_{BO}}
\newcommand{\htr}{\widehat{\mathfrak{h}}}
\newcommand{\htrtwo}{\widehat{\mathfrak{h}}^{(2)}}
\newcommand{\pop}{\hat{p}}
\newcommand{\qop}{\hat{q}}
\newcommand{\yop}{\hat{y}}
\newcommand{\deriv}[2]{\partial^{#1}_{#2}}
\newtheorem{theo}{Theorem}[section]

\newtheorem{lem}{Lemma}[section]
\theoremstyle{remark}
\newtheorem{rem}{Remark}[section]
\eqnobysec
\begin{document}
\title[Constrained quantum systems]{Semiclassical analysis of constrained quantum systems}
\author{G F Dell'Antonio\dag and L Tenuta\ddag}
\address{\dag Department of Mathematics, University of Rome I ``La Sapienza'', Piazzale Aldo Moro 2, 00185, Roma, Italy}
\address{\ddag Sissa/Isas - International School for Advanced Studies, Via Beirut 2-4, 34014, Trieste, Italy}
\eads{\mailto{gianfa@sissa.it}, \mailto{tenuta@sissa.it}}

\begin{abstract}
We study the dynamics of a quantum particle in $\field{R}{n+m}$ constrained by a strong potential force to stay within a distance of order
$\hbar$ (in suitable units) from a smooth $n-$dimensional submanifold $M$. We prove that in the semiclassical limit the evolution of the
wave function is approximated in norm, up to terms of order $\hbar^{1/2}$, by the evolution of a semiclassical wave packet centred on the
trajectory of the corresponding classical constrained system.
\end{abstract}

\section{Introduction}

The aim of this paper is to study the semiclassical limit of a nonrelativistic quantum Hamiltonian system in the configuration space
$\field{R}{n+m}$, constrained to a submanifold $M \subset \field{R}{n+m}$ by a confining potential which becomes infinite, in a suitable
sense
to be defined, when we move away from $M$.

We derive an effective Hamiltonian for the classical motion on $M$, using the technique developed in a series of papers by Hagedorn
(Hagedorn 1994, Hagedorn 1998 and references therein) to construct approximate solutions to the Schr\"odinger equation which are localized
along a classical trajectory.

We limit ourselves to Hamiltonians of the form
\begin{equation}\label{hamiltonianeps}
\eqalign{\Heps = \frac{|\pop|^{2}}{2} + V(\qop) + W^{\varepsilon}(\qop) \\
\pop := -i\hbar\nabla_{q} \qquad \qop := q\cdot}
\end{equation}
where $W^{\varepsilon}$ is the confining potential and $\varepsilon$ is a small parameter which we will make eventually go to zero (in
section \ref{couplingpotentials} we examine the motion of a particle in a magnetic field, which, under suitable conditions, can be put in
form \eref{hamiltonianeps}).

To explain the characteristic features of the method we employ, we first analyze in detail a number of explicit cases ($\field{R}{n}$
embedded into $\field{R}{n+m}$, a smooth curve embedded into a plane); we show then how the procedure generalizes to (non-flat) submanifolds
of arbitrary dimension and codimension.

The confining potential approach to imposing a constraint has been used often in the literature for a variety of reasons.

In classical mechanics, it has been employed mostly to ``realize holonomic constraints'' (Froese and Herbst 2001), \emph{i . e.}, to justify
the use of D'Alembert principle in deriving the Lagrange function for systems subject to holonomic time-independent constraints (which was
the starting point for the research performed in Takens (1980)). Other traditional applications include the analysis of magnetic traps and
mirrors, whose first complete mathematical discussion was given in Rubin and Ungar (1957), which was also the first rigorous investigation
in the field (a detailed treatment of these problems from the point of view of weak convergence, with extensions to arbitrary Riemannian
manifolds and molecular dynamics, can be found in Bornemann (1998)).

In quantum mechanics, the limit of large restoring force has been considered essentially for two reasons.

The first is that it offers a way,
different from the \emph{intrinsic} one (Henneaux and Teitelboim (1992) and references therein), to quantize constrained systems (da Costa
1981, da Costa 1982, Jensen and Koppe 1971, Kaplan \etal 1997, Maraner and Destri 1993, Mitchell 2001, Schuster and Jaffe 2003). 

The second is that in
\emph{mesoscopic physics} (\emph{i. e.}, the branch of physics which studies \emph{small} objects, like thin films and quantum wires) there
exist systems which have one, or more, dimensions much smaller than the others, and are then well described, in a zero order approximation,
by an $(n-k)$-dimensional confined system, $1\leq k \leq n-1$, (for the physical background and mathematical models see Duclos and Exner
(1995), Exner (2003) and
references therein). 

It was only recently (Froese and Herbst 2001, Teufel 2003) that a comparison between the classical and the quantum case was attempted.

The main problem one runs into is that, due to the Heisenberg principle, the mean value of the Hamiltonian operator diverges for every
initial
condition in the constraining limit (the better we localize the wave function on the submanifold $M$, the bigger the mean of the square of
the momentum becomes), while almost all theorems available in classical mechanics deal with finite energies.

To overcome these difficulties, Froese and Herbst state and prove a theorem on the classical case with unbounded energy, which, however,
does not seem very natural from a physical point of view, while in Teufel (2003) it is suggested to consider, instead of the limit of large
restoring forces, the limit of weak forces in the non-constraining directions (they are equivalent in classical mechanics, up to a
rescaling of space-time).

We propose a different approach, based on the fact that in quantum mechanics there exists an a priori length scale defined  through $\hbar$
(in units in which time and mass are of order one).

In real systems, like the mesoscopic ones mentioned above, the transversal directions contain at least some atoms, so any realistic layer
cannot become smaller than $\hbar$, which is the order of magnitude of atomic dimensions. Therefore, in our opinion, it is necessary to link
the squeezing scale, determined by the constraining potential, to the quantum scale given by $\hbar$. 

\subsection{A simple example}\label{simpleex}

To illustrate this point, we consider the standard two-dimensional example
\begin{equation}
\Heps = \frac{\pop^{2}_{x} + \pop^{2}_{y}}{2} + \frac{1}{2\varepsilon^{2}}\omega(x)^{2}y^{2}\label{standex},
\end{equation}
where $\omega: \field{R}{} \to \field{R}{}_{+}$ is an arbitrary smooth function which satisfies $\omega(x) \geq \omega_{*} > 0$ $\forall x
\in
\field{R}{}$. 

The squeezing scale is determined by $\varepsilon$, and we want it to be a function of $\hbar$, $\varepsilon = \varepsilon(\hbar)$. Since,
as we
argued before, $\varepsilon$ cannot become smaller than $\hbar$, and it has to go to zero when $\hbar \to 0$ (to achieve the constraining
limit), the simplest choice is 
\begin{equation}
\varepsilon = a\hbar^{\alpha} \qquad 0 < \alpha \leq 1 \qquad a \qquad {\rm fixed}> 0 
\end{equation}
(there is no loss of generality, since what matters is the behaviour of $\varepsilon(\hbar)$ when $\hbar \to 0$).

With this choice, the Hamiltonian \eref{standex} becomes
\begin{equation*}
\Hh = \frac{\pop^{2}_{x} + \pop^{2}_{y}}{2} + \frac{1}{2a^{2}\hbar^{2\alpha}}\omega(x)^{2}y^{2} 
\end{equation*}
and we want to examine the limiting behaviour of the dynamics generated by $\Hh$, when $\hbar \to 0$.

If we unitarily scale the transversal direction to factorize $\hbar$
\begin{equation*}
y \to \hbar^{(\alpha + 1)/2}y \qquad \deriv{}{y} \to \hbar^{-(\alpha + 1)/2}\deriv{}{y},
\end{equation*}
we get
\begin{equation*}
\Hh \to -\frac{\hbar^{2}}{2}\deriv{2}{x} + \hbar^{1 - \alpha}\Big[-\deriv{2}{y} + \frac{1}{2a^{2}}\omega(x)^{2}y^{2}\Big].
\end{equation*}
If $\alpha \neq 1$, using the same techniques illustrated in the next sections, it can be shown that the influence of the normal motion on
the longitudinal one is suppressed, and the effective Hamiltonian is the free one. Therefore, in the following, we consider only the more
interesting case $\alpha = 1$.

\subsection{Outline of the paper and summary of results}

In the next section we analyze a generalization of \eref{standex}, studying the case of a potential confining to a flat submanifold $M$ of
$\field{R}{n+m}$. We realize the constraining limit through dilations in the direction normal to $M$, \emph{i. e.}, we put
$W^{\varepsilon}(x, y) = W(x, y/\varepsilon)$. This allows us to consider generic dependence on the transversal variables, unlike
what is usually made in the literature (Bornemann 1998, Froese and Herbst 2001, Takens 1980), where the first non zero term in the Taylor
expansion of the potential around the constraint is the quadratic one, and so the problem is reduced to the analysis of harmonic motions. 

In section \ref{couplingpotentials} we consider a two-dimensional example where the constraining limit is realized through the more
traditional method of scaling of the coupling constant, \emph{i. e.}, $W^{\varepsilon}(x, y) = \varepsilon^{-2}W(x, y)$. In the case of a
\emph{spectrally smooth} potential confining to a \emph{nondegenerate critical curve} (for the definitions, see \ref{nondegsub} and
\ref{specsmooth}) the semiclassical limit motion we get along $M$ is the same as the homogenized classical motion found by Bornemann (1998).

In section \ref{genericdim}, we show that an analogous result holds for an $n$-dimensional nondegenerate critical submanifold embedded into
$\field{R}{n+m}$. We exploit Hagedorn's multiple scale technique to construct squeezed states whose centre and dispersion take account of
the (non-trivial) curved background.

Finally, we address an Hamiltonian showing the phenomenon of \emph{Takens chaos} (Bornemann 1998, Takens 1980), which is encountered
when the constraining potential is not spectrally smooth. In classical mechanics, the motion on the submanifold $M$ is not
deterministic anymore, \emph{i. e.}, it is not described by a natural mechanical system on $M$ and the limit set obtained forms a
\emph{funnel}. We show that the semiclassical limit offers a natural way to reduce (but however not to eliminate, in general) the
degeneracy, linking different trajectories in the funnel to different quantum initial conditions.    

\section{Constraints by normal dilations}

Let $M = \field{R}{n}$ and $W^{\varepsilon}(q) = W(x, y/\varepsilon)$, where we split $q \in \field{R}{n+m}$ as $(x, y)$, $x \in
\field{R}{n}$, $y \in \field{R}{m}$.

We suppose that 
\numparts
\begin{eqnarray}
V, W \in L^{2}_{{\rm loc}} \qquad \textrm{and are bounded from below}, \label{w1} \\
\lim_{|y| \to \infty} W(x, y) = \infty \qquad \forall x \in \field{R}{m}\qquad \textrm{\emph{(confining hypothesis)}}\label{w2}.
\end{eqnarray}  
\endnumparts

We impose also an implicit smoothness hypothesis on the potentials, through a condition on the resolvent of the reduced Hamiltonian
$\htr(x)$, to be defined below. 

As argued above, we put $\varepsilon = a\hbar$. Actually, since we have several normal directions, we can choose different
$\varepsilon/\hbar$ ratios  for each one.

Defining
\begin{equation}
y_{a} := \left( \begin{array}{ccc} \frac{y_{1}}{a_{1}} & \dots & \frac{y_{m}}{a_{m}} \end{array} \right )
\end{equation}
equation \eref{hamiltonian} becomes
\begin{equation}
\Hh =  \frac{|\pop|^{2}}{2} + V(x) + W(x, y_{a}/\hbar),
\end{equation}
where, for the sake of simplicity, we suppose that $V(q)$ does not depend on $y$.

Scaling the transversal directions by the dilation operator
\begin{equation}
(D_{\gamma}\psi)(x, y) = \gamma^{m/2}\psi(x, \gamma y),
\end{equation}
we get an Hamiltonian of the same form as the Born-Oppenheimer operator, used in molecular physics,
\begin{equation}
\eqalign{D^{\dagger}_{\hbar^{-1}}\Hh D_{\hbar^{-1}} =: \Hbo = -\frac{\hbar^{2}}{2}\Delta_{x} + \htr(x),\\
\htr(x) = -\frac{1}{2}\Delta_{y} + W(x, y_{a}) + V(x).}
\end{equation}

It follows from \eref{w1}, \eref{w2} that $\htr(x)$ is, for each $x$, a well defined self-adjoint operator, with compact resolvent and
nondegenerate ground state (Reed and Simon 1978).

We suppose in addition that $\htr(x)$ has a smooth dependence on $x$, namely that $(\htr(x) - \rmi)^{-1}$ is a $C^{l}$ function of $x$, for
some $l \geq 2$. This makes its eigenvalues $E(x)$ (which we will call also ``transversal'' or ``normal'' energy levels) $C^{l}$ functions
of $x$ away from crossings or absorption in the continuous spectrum.

The behaviour of Born-Oppenheimer Hamiltonian when $\hbar \to 0$ is well understood (Hagedorn 1994, Teufel 2003).

The transversal motion adiabatically decouples from the longitudinal one and stays approximately in a bound state of $\htr(x)$ for a fixed
value of $x$. On the other hand, the longitudinal motion depends on the transversal one because it feels an effective
potential which is equal to the normal energy. 

Using standard results (Hagedorn 1994) we can elaborate on this qualitative argument:
\begin{theo}\label{hagedorn} Suppose that there exists an open set $U \subset \field{R}{n}$ such that $\htr(x)$ has a nondegenerate
eigenvalue $E(x)$ for $x \in U$, with corresponding real normalized $C^{l}$ eigenfunction $\Phi(x)$. 

Let $a(t)$ and $\eta(t)$ be the solutions of the classical equations of motion with potential $E(x)$ (which exist and are unique since
$E(x)$ is $C^{l}(U)$ and bounded from below)
\begin{eqnarray}
\dot{a}(t) = \eta(t) \\
\dot{\eta}(t) = - \nabla E(a(t)),\\
a(0) = a_{0} \qquad \eta(0) = \eta_{0},
\end{eqnarray}
then, for $t \in [0, T]$,
\begin{eqnarray}\label{hagedornestimate}
\fl \Bigg\Arrowvert\exp\Big(-\frac{it}{\hbar}\Hh\Big)\varphi_{k}(A(0), B(0), \hbar, a(0), \eta(0), x)F(x)D_{\hbar^{-1}}\Phi(x) \nonumber\\
- \exp\Big(i\frac{S(t)}{\hbar}\Big)\varphi_{k}(A(t), B(t), \hbar, a(t), \eta(t),
x)F(x)D_{\hbar^{-1}}\Phi(x)\Bigg\Arrowvert_{L^{2}(\field{R}{n+m})} \nonumber\\
\lo= \Or(\hbar^{1/2}),
\end{eqnarray}
where $S(t)$ is the classical action, $A(t)$ and $B(t)$ are linked to the dispersions of $\varphi_{k}$ in (respectively) position and
momentum and $F$ is a cut function which is zero outside a neighbourhood of the classical trajectory $\{a(t): t\in [0, T]\}$.
\end{theo}
\begin{rem} The functions $\varphi_{k}(A, B, \hbar, a, \eta, x)$ were introduced by Hagedorn, to whom we refer for the notation
(Hagedorn 1998). They are a useful tool in studying the semiclassical limit of quantum mechanics and they coincide with the ``squeezed
states'' widely used in quantum optics (Combescure 1992). Essentially, they are minimal uncertainty wave packets with different spreads in
position and momentum. \end{rem}

\begin{rem} We will give a proof of a slightly more general version of theorem \eref{hagedorn} in sections \ref{couplingpotentials} and
\ref{genericdim}, where we analyze the Laplace-Beltrami operator in a curved space.
\end{rem}

\subsection{Comments and examples}

Let us analyze in greater detail the approximate evolution found in \eref{hagedornestimate}.

The transversal wave function $D_{\hbar^{-1}}\Phi(x)$ clearly describes a motion confined to the submanifold $M = \field{R}{n}$, since
\begin{equation}\label{meantransversal}
\eqalign{<\yop> = <D_{\hbar^{-1}}\Phi(x), yD_{\hbar^{-1}}\Phi(x)> = \hbar<\Phi(x), y\Phi(x)> = \Or(\hbar) \\
\fl (\Delta \yop_{i})^{2} = <D_{\hbar^{-1}}\Phi(x), y_{i}^{2}D_{\hbar^{-1}}\Phi(x)> - <D_{\hbar^{-1}}\Phi(x),
y_{i}D_{\hbar^{-1}}\Phi(x)>^{2} = \Or(\hbar^{2}),}
\end{equation}
while both $<\pop_{y}>$ and $<\Delta \pop_{y}>$ are $\Or(1)$.

One should note however that we did \emph{not} require $W$ to have a strict minimum on $M$. Actually this is not needed, since in our scale
the average position of the normal motion is always ``seen'' to be approximately zero, as equation \eref{meantransversal} shows. 

In the standard case where
\begin{equation}
W(x, y) =\frac{1}{2} \sum_{i=1}^{m} \omega_{i}(x)^{2}y_{i}^{2},
\end{equation}
the effective potential for the motion on $M$ will be
\begin{equation}\label{effectpot}
\eqalign{E_{n}(x) = \sum_{i=1}^{m} \frac{(n_{i} + 1/2)}{a_{i}}\omega_{i}(x) + V(x) = \sum_{i=1}^{m} \vartheta_{i}\omega_{i}(x) + V(x),\\
n := (n_{1}, \dots, n_{m}) \qquad \vartheta_{i} := \frac{(n_{i} + 1/2)}{a_{i}}.} 
\end{equation}

This is exactly the homogenized potential found by Bornemann (1998) and Takens (1980), where the $\vartheta_{i}$ are, in the classical case,
the adiabatic invariants associated to the normal oscillations (\emph{i. e.}, the energy-frequency ratios).

Varying the squeezing factors $a_{i}$, or the transversal wave function $\Phi(x)$, $\vartheta_{i}$ can be made to assume every positive
value (the value $\vartheta_{i} = 0$ can be obtained suppressing the $i$th mode as we explained in section \ref{simpleex}). The harmonic
potential is particular in this respect, because, as far as the effective potential is concerned, all normal states are equivalent, since
the various choices for $\Phi(x)$ correspond simply to suitable scalings of $\varepsilon$ and $\hbar$.

One could even use an $x$-dependent scale, $\varepsilon = a(x)\hbar$, without altering substantially the structure of equation
\eref{effectpot}.

Such a simple picture cannot be expected when $W$ is not harmonic.

In general, the effective potential will have a non-trivial dependence both on the parameters $\bi{a} := (a_{1}, \dots, a_{m})$ and the
transversal wave function. This gives a host of well-defined classical motions on $M$, whose form, however, cannot be given explicitly as
in the harmonic case.

It would be interesting, for instance, to compare the semiclassical effective Hamiltonians produced by a ``flat'' confining potential, like
the sextic harmonic oscillator,
\begin{equation}\label{sexticosc}
W(x, y) = V_{4}(x)y^{4} + V_{6}(x)y^{6}  \qquad (x, y)\in\field{R}{2} \qquad V_{6}(x) \geq V_{*}>0
\end{equation}  
with the corresponding homogenized classical motions (if any exists), to see if it is possible to reproduce them in a purely classical way.  

Unfortunately, the spectrum of the reduced Hamiltonian associated to \eref{sexticosc} is known only for particular values of the squeezing
parameter $a$. For example, if $a=1$ and $V_{4}(x)^{2} = 12V_{6}(x)^{3/2}$ it is known (Sk\'ala \etal 1996, Ushveridze 1994) that the
ground state is
\begin{equation} 
E_{0}(x) = \frac{V_{4}(x)}{2V_{6}(x)^{1/2}},
\end{equation}
but it is not possible to write an explicit expression for all values of $a$.

\section{Constraints by scaling of coupling constant: a curve in a plane}\label{couplingpotentials}

In this section we analyze, in a fairly detailed way, a two-dimensional example where $W^{\varepsilon} = \varepsilon^{-2}W$. It allows to
explain the main differences between the curved and the flat case, avoiding technical complications arising from higher codimensions, which
are not essential for the result, and will be illustrated in next section.

We suppose, in the same spirit of \eref{w1}, that $V$ and $W$ are $C^{\infty}$ and non-negative, but, as is customary in classical
mechanics (Bornemann 1998, Takens 1980), we replace \eref{w2} with the hypothesis that $W$ is a spectrally smooth potential constraining to
a nondegenerate critical curve $M$(\ref{nondegsub} and \ref{specsmooth}). 

Our starting Hamiltonian (with the prescription $\varepsilon = a\hbar$) will be then
\begin{equation}\label{hamiltonian}
\Hh = \frac{\pop_{x}^{2} + \pop_{y}^{2}}{2} + V(x, y) + (a\hbar)^{-2}W(x, y).
\end{equation}

Squeezed states are particularly suited to studying this sort of situations, where $M$ is not flat, because, as \eref{hagedornestimate}
shows, the evolution of a localized state is approximately described (for a bounded time interval) by localized states. This allows
us to analyze the motion using one coordinate chart only and therefore local expressions for the operators involved. 

Essentially, what we will do here is to adapt the arguments of the last section to a curved case, constructing an approximate solution to
the Schr\"odinger equation which, in suitable coordinates, is still given by a squezeed state in the longitudinal direction and an
(harmonic) oscillation in the transversal one.

\subsection{The Hamiltonian in curvilinear coordinates}

We fix a tubular neighbourhood $\mathscr{V}$ of $M$, and we consider a single chart of tubular coordinates, defined on $\mathscr{U}
\subset \mathscr{V}$.

This simply means that, given a local parametric representation of $M$ in terms of its arc length $s$, $q_{M}(s) = (x_{M}(s), y_{M}(s))$, we
can write (for $q \in \mathscr{U}$)
\begin{equation}\label{tubcoord}
q(s, u) = q_{M}(s) + u\bi{n}(s),
\end{equation}
where $\bi{n}(s)$ is the unit normal of $M$.

In writing \eref{tubcoord} we used the natural linear structure of tubular coordinates. A more invariant, but less manageable, relation
would be
\begin{equation*}
q = \exp_{q_{M}}q^{\perp} \qquad q_{M} \in M \qquad q^{\perp} \in T_{q_{M}}M^{\perp},
\end{equation*}
where $\exp$ is the geodesic exponential map (Lang 1995). In the following, however, we will stick to \eref{tubcoord}.

When $q$ varies over $\mathscr{U}$, $s$ and $u$ vary, respectively, over two intervals $I$ and $J$.

\begin{lem} The Hilbert space $L^{2}(\mathscr{U}, dq)$ is isometric to $L^{2}(I\times J, dsdu)$. \end{lem}
\begin{proof} This well-known lemma results from two facts.

First, the choice of curvilinear coordinates provides an isometry of $L^{2}(\mathscr{U}, dq)$ to $L^{2}(I\times J, g^{1/2}dsdu)$, where 
\begin{equation}
g^{1/2} = 1 - k(s)u
\end{equation}
is the Jacobian of the transformation $(x, y) \to (s, u)$, and $k(s)$ is the curvature of $M$.

Second, the multiplication by $g^{1/4}$ is a unitary operator from $L^{2}(I\times J, g^{1/2}dsdu)$ to $L^{2}(I\times J, dsdu)$.
\end{proof}
In the following, we will denote the isometry constructed above by $\widehat{U}: L^{2}(\mathscr{U}, dq) \to L^{2}(I\times J, dsdu)$.

We remark that $\widehat{U}$ maps $\cinf(\mathscr{U})$ onto $\cinf(I\times J)$ and $\Hh$ maps $\cinf(\mathscr{U})$ into
$\cinf(\mathscr{U})$, so, denoting, with abuse of notation, the restrictions of $\widehat{U}$ and $\Hh$ to  $\cinf$ functions with the same
symbols, we have 
\begin{eqnarray}
\widehat{U}\Hh\widehat{U}^{\dagger}: \cinf(I\times J) \to \cinf(I\times J) \nonumber \\
\fl \widehat{U}\Hh\widehat{U}^{\dagger} = -\frac{\hbar^{2}}{2}\frac{1}{(1 - k(s)u)^{1/2}}\deriv{}{s}\Big(\frac{1}{1 -
k(s)u}\deriv{}{s}\frac{\cdot}{(1 - k(s)u)^{1/2}}\Big) \nonumber \\
- \frac{\hbar^{2}}{2}\deriv{2}{u} - \frac{\hbar^{2}}{8} \frac{k(s)^{2}}{(1 - k(s)u)^{2}} + \tilde{V}(s, u) + (a\hbar)^{-2}\tilde{W}(s, u)\\
\lo= -\frac{\hbar^{2}}{2(1 - k(s)u)^{2}}\deriv{2}{s} - \frac{\hbar^{2}\dot{k}(s)u}{(1 - k(s)u)^{3}}\deriv{}{s} - \hbar^{2}Q(s, u)\\
 - \frac{\hbar^{2}}{2}\deriv{2}{u} + \tilde{V}(s, u) + (a\hbar)^{-2}\tilde{W}(s, u),
\end{eqnarray}
where $\tilde{V}$ and $\tilde{W}$ are $V$ and $W$ written in curvilinear coordinates and $\hbar^{2}Q$ is an extrapotential of purely quantum
origin which depends on the curvature $k(s)$ (da Costa 1981, Jensen and Koppe 1971). It appears also in mesoscopic physics, and can give
rise to interesting phenomena, like bound states, in a quantum waveguide (Duclos and Exner 1995). However, it will not concern us, since it
disappears in the lowest order of semiclassical approximation. 

Using again a dilation operator in the transversal direction $u$,
\begin{eqnarray}
D_{\gamma}: L^{2}(I\times J_{\gamma}, dsdu) \to L^{2}(I\times J, dsdu) \nonumber\\
(D_{\gamma}\psi)(s, u) = \gamma^{1/2}\psi(s, \gamma u) \nonumber \\
J_{\gamma}:= \{\gamma u: u\in J\},
\end{eqnarray}
we get the final form of the Hamiltonian which we will employ in the estimates:
\begin{eqnarray} \label{localham}
\Hbo : \cinf(I\times J_{\hbar^{-1}}) \to \cinf(I\times J_{\hbar^{-1}}) \nonumber\\
\fl \Hbo := D^{\dagger}_{\hbar^{-1}}\widehat{U}\Hh\widehat{U}^{\dagger}D_{\hbar^{-1}} \nonumber \\
\lo= - \frac{\hbar^{2}}{2(1 - \hbar k(s)u)^{2}}\deriv{2}{s} - \frac{\hbar^{3}\dot{k}(s)u}{(1 - \hbar k(s)u)^{3}}\deriv{}{s} - \hbar^{2}Q(s,
\hbar u) + \htr(s),
\end{eqnarray}
where
\begin{equation}
\htr(s) =  - \frac{1}{2}\deriv{2}{u} + \tilde{V}(s, \hbar u) + (a\hbar)^{-2}\tilde{W}(s, \hbar u).
\end{equation}

\begin{rem} Note that 
\begin{eqnarray}
\fl (a\hbar)^{-2}\tilde{W}(s, \hbar u) = \frac{1}{2a^{2}}\deriv{2}{u}\tilde{W}(s, 0)u^{2} + \frac{\hbar}{6a^{2}}\deriv{3}{u}\tilde{W}(s,
0)u^{3} + \frac{1}{6a^{2}\hbar^{2}}\int_{0}^{\hbar u} dv (\hbar u -v)^{3}\deriv{4}{u}\tilde{W}(s, v) \nonumber \\
\lo= \frac{1}{2a^{2}}\omega(s)^{2}u^{2} +  \frac{\hbar}{6a^{2}}\deriv{3}{u}\tilde{W}(s, 0)u^{3} + R_{3}(\hbar, u) , \\
\fl \tilde{V}(s, \hbar u) = \tilde{V}(s, 0) + \hbar u\deriv{}{u}\tilde{V}(s, 0) + \int_{0}^{\hbar u} dv (\hbar u -
v)\deriv{2}{u}\tilde{V}(s, v) \nonumber \\
\lo= \tilde{V}(s, 0) + \hbar u\deriv{}{u}\tilde{V}(s, 0) + R_{1}(\hbar, u).
\end{eqnarray}

The scaling in the normal direction eliminates the dependence of $\htr$ on $\hbar$ only at the lowest order in the Taylor expansion around
the constraint (which is the quadratic one since $M$ is a nondegenerate critical curve).

Frow now on, we will denote by $\htrtwo(s)$ the harmonic part of $\htr(s)$:
\begin{equation}
\htrtwo(s):= -\frac{1}{2}\deriv{2}{u} + \frac{1}{2a^{2}}\omega(s)^{2}u^{2} + \tilde{V}(s, 0).
\end{equation}
 \end{rem}

\subsection{The approximate evolution}

In this subsection we prove the
\begin{theo}\label{curvedcase} Let $\Phi(s, u)$ be a real normalized eigenstate of $\htrtwo(s)$, considered as an operator on
$L^{2}(\field{R}{}, du)$, with eigenvalue $E(s)$. Let $a(t)$ and $\eta(t)$ be the solutions of the classical equations of motion with
potential $E(s)$, and let $F(s, v)$ be a function in $\cinf(I\times J)$ which is equal to $1$ for $s$ in a neighbourhood of the trajectory
$\{a(t): t \in [0, T]\}$ and $v$ near to $0$. 

Then
\begin{eqnarray}
\fl \Big\Arrowvert \exp\Big(-\frac{it}{\hbar}\Hh\Big) \widehat{U}^{\dagger}D_{\hbar^{-1}}\varphi_{k}(A(0), B(0), \hbar, a(0), \eta(0),
s)F(s, \hbar u)\Phi(s, u) \nonumber \\
- \exp\Big(\frac{iS(t)}{\hbar}\Big)\widehat{U}^{\dagger}D_{\hbar^{-1}}\varphi_{k}(A(t), B(t), \hbar, a(t), \eta(t), s)F(s, \hbar u)\Phi(s,
u)\Big\Arrowvert \nonumber \\
\lo= \Or(\hbar^{1/2}),
\end{eqnarray}
where $S(t)$ is the classical action associated to $(a(t), \eta(t))$.
\end{theo}

\begin{rem} The function $\varphi_{k}(A(t), B(t), \hbar, a(t), \eta(t), s)F(s, \hbar u)\Phi(s, u)$ is in $\cinf(I\times J_{\hbar^{-1}})$, so
$\widehat{U}^{\dagger}D_{\hbar^{-1}}\varphi_{k}(A(t), B(t), \hbar, a(t), \eta(t), s)F(s, \hbar u)\Phi(s, u)$ belongs to
$\cinf(\mathscr{U})$.
\end{rem}

The proof will follow closely the pattern developed by Hagedorn (Hagedorn 1994), but the remainder we get is different from that found by
him, since $\htr$ contains terms of order $\hbar$ and the kinetic part of \eref{localham} is not simply $- (\hbar^{2}/2)\deriv{2}{s}$.

The basic tool we use is a simple application of the fundamental theorem of calculus (also known as Duhamel formula). We give it without
proof.
\begin{lem}\label{duhamel} Suppose $\Hh$ is a family of self-adjoint operators for $\hbar > 0$. Suppose $\psi(\hbar, t)$ belongs to the
domain of $\Hh$, is continuously differentiable in $t$, and approximately solves the Schr\"odinger equation in the sense that
\begin{equation}
\rmi\hbar\deriv{}{t}\psi(\hbar, t) = \Hh\psi(\hbar, t) + \zeta(\hbar, t),
\end{equation}
where $\zeta(\hbar, t)$ satisfies 
\begin{equation}
||\zeta(\hbar, t)|| \leq \mu(\hbar, t)
\end{equation}
for $0 \leq t \leq T$. Suppose $\Psi(\hbar, t)$ is the exact solution to the equation
\begin{equation}
\rmi\hbar\deriv{}{t}\Psi(\hbar, t) = \Hh\Psi(\hbar, t)
\end{equation}
with initial condition $\Psi(\hbar, 0) = \psi(\hbar, 0)$. 

Then, for $0 \leq t \leq T$, we have
\begin{equation}
||\Psi(\hbar, t) - \psi(\hbar, t)|| \leq \hbar^{-1} \int_{0}^{T} d\tau \mu(\hbar, \tau) .
\end{equation}
\end{lem}

Suppose now that $\psi_{\textrm{ap}}(s, u, t) \in \cinf(I\times J_{\hbar^{-1}})$ is an approximate solution to the Schr\"odinger equation
associated to the local Hamiltonian \eref{localham},
\begin{equation}\label{localschro}
\rmi\hbar\deriv{}{t}\psi_{\textrm{ap}} = \Hbo\psi_{\textrm{ap}} + \zeta(\hbar, t)
\end{equation}
with 
\begin{equation}\label{curvestim}
||\zeta(\hbar, t)||_{L^{2}(I\times J_{\hbar^{-1}})} = \Or(\hbar^{3/2}) \qquad \textrm{for} \qquad 0 \leq t \leq T.
\end{equation}

This implies that
\begin{equation*}
\rmi\hbar\deriv{}{t}\widehat{U}^{\dagger}D_{\hbar^{-1}}\psi_{\textrm{ap}} = \Hh\widehat{U}^{\dagger}D_{\hbar^{-1}}\psi_{\textrm{ap}} +
\tilde{\zeta}(\hbar, t),
\end{equation*}
with $||\tilde{\zeta}(\hbar, t)||_{L^{2}(\mathscr{U})} = \Or(\hbar^{3/2})$.

Using lemma \eref{duhamel} we get finally
\begin{equation*}
\Big\Arrowvert\exp\Big(-\frac{it}{\hbar}\Hh\Big)\widehat{U}^{\dagger}D_{\hbar^{-1}}\psi_{\textrm{ap}}(t=0) -
\widehat{U}^{\dagger}D_{\hbar^{-1}}\psi_{\textrm{ap}}(t)\Big\Arrowvert_{L^{2}(\field{R}{2})} = \Or(\hbar^{1/2}).
\end{equation*}

Therefore, to prove theorem \eref{curvedcase} we will construct an approximate solution to \eref{localschro}, of the form
\begin{equation}
\psi_{\textrm{ap}}(s, u, t) = \psi_{0}(s, u, t) + \hbar\psi_{2}^{\perp}(s, u, t),
\end{equation}
with $\psi_{0}(s, u, t) = \exp(\rmi S(t)/\hbar)\varphi_{k}(A(t), B(t), \hbar, a(t), \eta(t), s)F(s, \hbar u)\Phi(s, u)$ (the notation
$\psi_{2}^{\perp}$ means that the transversal part of this term is orthogonal to $\Phi$). 

An educated guess about the form of the remainder $\psi_{2}^{\perp}$ can be made employing a multiple scale technique, which allows to split
the adiabatic and the semiclassical scale. 

We will elaborate on this procedure in the more complicated case of next section, so here we limit ourselves to verify
that the right choice is
\begin{eqnarray}
\fl \psi_{2}^{\perp}(s, u, t) = \varphi_{k}(A(t), B(t), \hbar, a(t), \eta(t), s)F(s, \hbar u) \nonumber  \\
\times \hat{r}(s)\Big[\rmi\eta(t)\deriv{}{s}\Phi(s, u) - \eta(t)^{2}k(s)u\Phi(s, u) - \deriv{}{u}\tilde{V}(s, 0)u\Phi(s, u) \nonumber \\
-\frac{1}{6a^{2}}\deriv{3}{u}\tilde{W}(s, 0)u^{3}\Phi(s, u)\Big],
\end{eqnarray} 
where $\hat{r}(s)$ is the bounded inverse of the restriction of $[\htr^{(2)}(s) - E(s)]$ to the orthogonal complement of
$\Phi(s, u)$ in $L^{2}(\field{R}{}, du)$.

Estimate \eref{curvestim} will follow if we note the following facts:
\begin{enumerate}
\item \emph{the terms containing derivatives of $F$ are $\Or(\hbar^{\infty})$}. For instance,
\begin{eqnarray}
\fl \int_{I\times J_{\hbar^{-1}}} dsdu |\deriv{}{u}F(s, \hbar u)\varphi_{k}(s)\deriv{}{u}\Phi(s, u)|^{2} \\
\lo= \int_{I\times J} dsdv |\deriv{}{v}F(s, v)\varphi_{k}(s)\hbar^{1/2}\deriv{}{u}\Phi(s, v\hbar^{-1})|^{2} < \exp(- C\hbar^{-1}) \nonumber, 
\end{eqnarray} 
since $\deriv{}{v}F$ has support away from zero in $v$, and $\deriv{}{u}\Phi(s, v\hbar^{-1})$ is a polinomial times a Gaussian, in $u =
v\hbar^{-1}$. 

The derivatives with respect to $s$ can be estimated in the same way, since $\varphi_{k}$ is a Gaussian in $[s - a(t)]/\hbar^{1/2}$.

\item The term 
\begin{equation*}
\frac{\hbar^{3}\dot{k}(s)u}{(1 - \hbar k(s)u)^{3}}\deriv{}{s}\psi_{\textrm{ap}}
\end{equation*}
is $\Or(\hbar^{2})$ since $\deriv{}{s}\varphi_{k}$ is $\Or(\hbar^{-1})$.

\item The term
\begin{equation*}
\hbar^{2}Q(s, \hbar u)\psi_{\textrm{ap}}
\end{equation*}
is $\Or(\hbar^{2})$ since $Q(s, u)$ is bounded on the support of $F$.

\item The last term is
\begin{eqnarray}
\fl \htr(s)\psi_{\textrm{ap}} = \htrtwo\psi_{\textrm{ap}} + \frac{\hbar}{6a^{2}}\deriv{3}{u}\tilde{W}(s, 0)u^{3}\psi_{0} + \hbar
u\deriv{}{u}\tilde{V}(s, 0)\psi_{0} + R_{3}(\hbar, u)\psi_{\textrm{ap}} \nonumber \\
+ R_{1}(\hbar u)\psi_{\textrm{ap}} + \Or(\hbar^{2}) = E(s)\psi_{\textrm{ap}} + \rmi\hbar\varphi_{k}(s)F(s, \hbar u)\eta(t)\deriv{}{s}\Phi(s,
u) \nonumber  \\
+ \hbar\eta(t)^{2}k(s)u\psi_{0} + \Or(\hbar^{2}) \nonumber,
\end{eqnarray}
since $R_{3}(\hbar, u)$ and $R_{1}(\hbar, u)$ are $\Or(\hbar^{2})$ on the support of $F$.

\item The terms left combine themselves with the kinetic part and the time derivative of $\psi_{\textrm{ap}}$ to give \eref{curvestim}.
\end{enumerate}

\begin{rem} 
The effective motion on $M$ is given by the potential
\begin{equation}
E_{n}(s) = \frac{(n + 1/2)}{a}\omega(s) + \tilde{V}(s, 0) = \vartheta\omega(s) + \tilde{V}(s, 0),
\end{equation}
and is equal, also in this case, to the homogenized classical motion.

The hypotheses that $M$ is a nondegenerate critical curve and $W$ is spectrally smooth imply that the normal oscillation is harmonic, and so
all transversal states are equivalent. \end{rem}

\subsection{The magnetic trap}

Using theorem \eref{curvedcase} we can analyze the dynamics of a nonrelativistic particle in a strong magnetic field (magnetic trap).

We suppose that the field is ``strongly axially symmetric'', \emph{i. e.}, that the vector potential is given, in cylindrical coordinates,
by
\begin{equation} 
\bi{A}(r, z) = \mathscr{A}(r, z)\btheta.
\end{equation}

The Hamiltonian is 
\begin{equation}\label{magneticham}
\widehat{H} = \frac{1}{2m}\Big(\pop - \frac{e}{c}\bi{A}\Big)^{2}.
\end{equation}

Since $\textrm{div}\bi{A} = 0$, in the susbspace with zero angular momentum in the $z$ direction \eref{magneticham} becomes 
\begin{equation*}
\widehat{H}^{0} = -\frac{\hbar^{2}}{2m}\frac{1}{r}\deriv{}{r}(r\deriv{}{r}) - \frac{\hbar^{2}}{2m}\deriv{2}{z} +
\frac{e^{2}}{2mc^{2}}\mathscr{A}(r, z)^{2},
\end{equation*}
or, scaling the wave function by the isometry
\begin{eqnarray}
\widehat{V}: L^{2}(\field{R}{}_{+}\times\field{R}{}, rdrdz) \to L^{2}(\field{R}{}_{+}\times\field{R}{}, drdz) \nonumber \\
\widehat{V}\psi = r^{1/2}\psi,
\end{eqnarray}
\begin{equation}
\widehat{V}\widehat{H}^{0}\widehat{V}^{\dagger} = -\frac{\hbar^{2}}{2m}\deriv{2}{r} - \frac{\hbar^{2}}{8mr^{2}} -
\frac{\hbar^{2}}{2m}\deriv{2}{z} + \frac{e^{2}}{2mc^{2}}\mathscr{A}(r, z)^{2}.
\end{equation}

If we put $m = 1$ and consider the case of large electric charge, $c/e = a\hbar$, we get in the end
\begin{equation}
\Hh:= -\frac{\hbar^{2}}{2}\deriv{2}{r} - \frac{\hbar^{2}}{2}\deriv{2}{z} - \frac{\hbar^{2}}{8r^{2}} +
\frac{1}{2a^{2}\hbar^{2}}\mathscr{A}(r, z)^{2},
\end{equation}
which, except for the centrifugal term, is of the form \eref{hamiltonian}, with $W(r, z) = \mathscr{A}(r, z)^{2}/2$.

Theorem \eref{curvedcase} tells us that, if we consider an initial state localized away from the origin, the semiclassical motion is
constrained along the curve $\mathscr{A}(r, z) = 0$, with effective potential given by
\begin{equation}
E(s) = \vartheta\{\deriv{2}{u}[\tilde{\mathscr{A}}(s, u)]^{2}/2\}^{1/2}_{|_{u=0}} = \vartheta|\deriv{}{u}\tilde{\mathscr{A}}(s, 0)| =
\vartheta|\bi{B}(s, 0)|,
\end{equation}
where $\bi{B}$ is the magnetic field strength.

\section{Constraints by scaling of coupling constant: general case}\label{genericdim}

When the submanifold $M$ has dimension (and codimension) greater than one, the theory developed in foregoing sections has to be generalized
essentially in two aspects.

First, if $\dim M >1$, the metric $G_{M}$, induced by the Euclidean metric of $\field{R}{n+m}$ on $M$, may not be trivial, so both the
classical motion of the centre of the squeezed state and the evolution of the dispersion matrices $A$ and $B$ have to be modified to take
this into account. Thinking about the results we got above, it is not difficult to derive the new classical equations; we will simply
obtain a motion on a Riemannian manifold with metric $G_{M}(x)$ in the presence of a potential $E(x)$ which is an eigenvalue of the reduced
Hamiltonian. In local coordinates this means (see, for instance, Abraham and Marsden 1978)

\begin{eqnarray}\label{classicmultidim}
\dot{a}(t) = \eta(t) \\
\dot{\eta}(t) = -\Gamma(a(t))(\eta, \eta) - G_{M}^{-1}(a(t))\nabla_{x} E(a(t)) ,
\end{eqnarray}  
where $\Gamma(\eta, \eta)^{i} = \Gamma^{i}_{jk}\eta^{j}\eta^{k}$ ($\Gamma^{i}_{jk}$ are the Christoffel symbols associated to $G_{M}$) and
$\nabla_{x}$ denotes the column vector whose coordinates are $\deriv{}{i} :=\deriv{}{x_{i}}$.

The equations for the dispersion matrices are a bit more complicated, but, as we will see below, they can be derived, using the Hagedorn
multiple scale method, from the term of order $\hbar$ of the formal expansion of the solution of the Schr\"odinger equation in powers of
$\hbar^{1/2}$.

The second point is that, if $\textrm{codim}\, M >1$, the Euclidean metric written in tubular coordinates is not diagonal anymore.

In a formal expansion of the Hamiltonian $\Heps$ around the constraint, the off-diagonal terms give rise, as first noted by Maraner and
Destri (Maraner and Destri 1993; see also Froese and Herbst 2001, Mitchell 2001, Schuster and Jaffe 2003 and references therein) to an
induced gauge field which minimally couples the longitudinal and the transversal motion. This gauge field is linked to the normal
connection for the embedding $M \subset \field{R}{n+m}$ (see, for instance, Spivak 1979) and it certainly vanishes if $\textrm{codim}\,
M=1$.

At first sight, it might seem that in this case we can no longer split the motion into a tangential and a normal part, even in the
semiclassical limit.

Actually this is not true, since, applied to a squeezed state, the gauge coupling is of order $\hbar$, and, due to the antisymmetric
character of the normal fundamental form, it maps an eigenstate of the reduced Hamiltonian into a state which is ortogonal to it.

According to the proof of theorem \eref{curvedcase}, this means that, if we start from an initial state which is concentrated along a
classical trajectory, and we study its evolution when $\hbar$ goes to zero, the gauge term contributes only to the remainder and not to the
leading term of the expansion in powers of $\hbar^{1/2}$, which is again given by a wave packet in the longitudinal variables times an
eigenstate of the normal Hamiltonian.  

In principle, higher order corrections can be calculated following the procedure developed by Hagedorn (Hagedorn 1994), even though in the
general case the formulae can be cumbersome.

In the following we will give some details of the calculations that justify these claims, even though, given the previous warnings, they are
analogous to those of the two-dimensional case.

\subsection{The Hamiltonian in tubular coordinates}

By the tubular neighbourhood theorem (Lang 1995), given a local chart $\zeta^{-1}: E \subset M \to \field{R}{n}$ for the submanifold $M$,
and a $\delta$ small enough, there exists a diffeomorphism between 
\begin{eqnarray}
\mathscr{E}(\delta) := \{q \in \field{R}{n+m}: d(q, E)<\delta\},\\
d(q, E) := \inf\{|q-e|: e \in E\} \nonumber
\end{eqnarray} 
and the open subset of the normal bundle of $M$ given by
\begin{equation}
TE_{\delta}^{\perp} := \{(e, n): e \in E, n \in T_{e}E^{\perp}, |n| < \delta\}.
\end{equation}
The diffeomorphism can be chosen to be
\begin{equation*}
f(e, n) = e + n ,
\end{equation*}
where we have identified every fibre $T_{e}E^{\perp}$ with a subspace of $\field{R}{n+m}$.

This means that, given a (local) basis for the normal bundle $\{n_{k}(e)\}_{k=1}^{m}$, we can write every point in $\mathscr{E}(\delta)$ as
\begin{equation}
q = \zeta(x) + y_{k}n_{k}(\zeta(x)) \qquad x \in \field{R}{n} \quad y \in \field{R}{m}
\end{equation}
(summation over repeated indices is understood).

Starting from the above expression, we can calculate the coordinate form of the basis for the tangent space in a point of
$\mathscr{E}(\delta)$ simply differentiating with respect to a coordinate $x_{i}$ or $y_{k}$, and then calculate the scalar product of two
basis elements to get the local form of the metric.

The result is
\begin{equation}\label{metric}
G(x, y) = \left( \begin{array}{cc} I & N \\
                                                   0 & I      \end{array} \right) \left( \begin{array}{cc} G_{M}(I - S)^{2} & 0 \\
                                                                                                                                        0  &
I \end{array} \right) \left( \begin{array}{cc} I & N \\
                                                                    0 & I \end{array} \right )^{T},
\end{equation}
where 
\begin{eqnarray}
N_{i, h}(x, y) = y_{k}\beta^{kh}_{i}(x) \qquad \beta^{kh}_{i} = n_{k}\cdot\deriv{}{i}n_{h}, \label{N} \\
S_{i, j}(x, y) = y_{k}(G_{M}^{-1})_{il}\alpha^{k}_{lj}(x) \qquad \alpha^{k}_{lj}(x) = n_{k} \cdot \deriv{}{l}t_{j} \label{S} ,
\end{eqnarray}
and $t_{l}$ denotes the basis for the tangent space (the index $k$ and $h$ always refer to the normal coordinates, while the other indices
refer to the tangential coordinates).

$\beta^{kh}_{i}$ and $\alpha^{k}_{il}$ are called, respectively, \emph{normal fundamental form} and \emph{second fundamental form} of the
submanifold $M$. Together with the metric $G_{M}$, they characterize completely the embedding of $M$ into $\field{R}{n+m}$, up to a
Euclidean motion (Spivak 1979). It is important to stress that $\beta^{hk}_{i} = - \beta^{kh}_{i}$, so, when $\textrm{codim}\, M =1$,
$\beta$ is identically zero.

Using \eref{metric}, we can write the Hamiltonian in tubular coordinates, but, as we did in the two-dimensional case, we have to modify the
volume form given by $g(x, y)^{1/2} := [\det G(x, y)]^{1/2}$, in order to get wave functions which have the right normalization when
integrated over the submanifold $M$. 

After this, we have to dilate the normal coordinates by $\hbar$, in order to separate the reduced Hamiltonian from the longitudinal part.

This can be achieved by the unitary operator
\begin{eqnarray}
(\widehat{V}\psi)(x, y) = \Bigg(\frac{g_{M}(x)}{g(x, y)}\Bigg)^{1/4}\hbar^{-m/2}\psi(x, y/\hbar) \\
\widehat{V}: L^{2}(\mathscr{E}(\delta/\hbar), g_{M}(x)^{1/2}dxdy) \to L^{2}(\mathscr{E}(\delta), g(x, y)^{1/2}dxdy) \nonumber
\end{eqnarray}
where $g_{M}(x) := \det G_{M}(x)$.

The result in the end is
\begin{eqnarray}\label{hbomultidim}
\fl \Hbo = \widehat{V}^{\dagger} \Hh \widehat{V} \nonumber \\
\lo{=} -\frac{\hbar^{2}}{2}\rho_{\hbar}(x, y)^{-1/4}g_{M}^{-1/2}
\left( \begin{array}{cc} \nabla_{x}^{T} - \nabla_{y}^{T} N^{T}(x, y), & \hbar^{-1}\nabla_{y}^{T} \end{array} \right)
g_{M}^{1/2}\rho_{\hbar}^{1/2} \nonumber \\
\cdot \left( \begin{array}{cc} [I - \hbar S(x, y)]^{-2} G_{M}^{-1}(x) & 0 \\
                                                       0 & I \end{array} \right)
\left( \begin{array}{c} \nabla_{x} - N(x, y)\nabla_{y} \\
                                     \hbar^{-1} \nabla_{y} \end{array} \right) \rho_{\hbar}^{-1/4} \nonumber \\
+ V(x + \hbar y) + (a\hbar)^{-2}W(x+\hbar y),
\end{eqnarray}
where 
\begin{equation}
\rho_{\hbar}(x, y) = \frac{g(x, \hbar y)}{g_{M}(x)} \qquad .
\end{equation}

When we further expand the equation \eref{hbomultidim}, the terms containing $\rho_{\hbar}(x, y)$ give rise to additive corrections which
depend only on the second derivatives (or the square of the first derivatives) of $\ln\rho_{\hbar}$. They are of order at least
$\hbar^{2}$. This can be understood if we note that $S(x, y)$ is linear in $y$, the second derivatives with respect to $x$ are multiplied by
$\hbar^{2}$ and
\begin{eqnarray}
\fl \ln\rho_{\hbar}(x, y) = \ln \frac{\det \{G_{M}[I -\hbar S(x, y)]^{2}\}}{\det G_{M}} = 2\ln\det(I - \hbar S(x, y)) = 2\Tr\ln(I - \hbar S)
\nonumber \\
\lo= -2\hbar \Tr(S) - \hbar^{2}\Tr(S^{2}) + \Or(\hbar^{3}) \qquad .
\end{eqnarray}
Therefore, in the following, we will put $\rho_{\hbar} = 1$ without other comments.

Expanding the potentials $V$ and $W$, we obtain the reduced Hamiltonian
\begin{equation}
\htrtwo(x) = -\frac{1}{2}\Delta_{y} + \frac{1}{2a^{2}}y^{T}H(x)y + V(x),
\end{equation}
where $H(x)$ is the matrix of the Hessian operator in the basis $\{n_{k}(\zeta(x))\}$. The hypothesis that $W$ has a smooth spectral
decomposition implies that we can choose the $n_{k}$ to be eigenvectors of $H$, so we can write $y^{T}H(x)y = \sum_{\lambda, k_{\lambda}}
\omega_{\lambda}^{2}(x) y_{\lambda, k_{\lambda}}^{2}$.

We will see in next subsection that, as before, the higher order terms in the Taylor expansion must be included in the remainder.

\subsection{The approximate evolution}

To construct approximate solutions to the Schr\"ordinger equation
\begin{equation}\label{schrodmultidim}
\rmi\hbar\deriv{}{t}\psi = \Hbo\psi
\end{equation}
we use the same procedure outlined in previous sections, which is based on the multiple scale expansion developed by Hagedorn (Hagedorn
1994). The operator \eref{hbomultidim} is not of the standard form studied in the literature, so we briefly explain the modifications
needed to cope with this case.

When all the terms have been spelled out, \eref{hbomultidim} has the form of an elliptic differential operator in $x$ and $y$, with
coefficients which depend on $x$ and $y$ as well as $\hbar$, plus the reduced Hamiltonian, plus a remainder of order $\hbar$, which comes
from the Taylor expansion of $V(x+\hbar y)$ and $W(x+\hbar y)$ up to first and third order, respectively.

According to the Hagedorn method, to split the adiabatic and the semiclassical effects, we have to introduce a fictitious new variable
\begin{equation}\label{xi}
\xi := \frac{x-a(t)}{\hbar^{1/2}},
\end{equation}
which measures the ``deviation'' of the quantum evolution from the classical one, and consider $\xi$ as an independent variable in the
formal manipulations. 

Associated to $\xi$, there is an auxiliary wave function, $\tilde{\psi}(x, y, \xi; t)$, which satisfies the equation obtained substituting 
\begin{equation*}
\tilde{\psi}\Big(x, y, \frac{x-a(t)}{\hbar^{1/2}}; t\Big) 
\end{equation*}
into \eref{schrodmultidim}, and adding to the right-hand side the term $E(a(t)+\hbar^{1/2}\xi) - E(x)$, which formally equals zero when $\xi
= [x-a(t)]/\hbar^{1/2}$, where $E(x)$ is a fixed eigenvalue of $\htrtwo(x)$, with multiplicity $1$.

When we perform this substitution, we replace the $x$ dependence in the coefficients of the differential terms with a dependence on $a(t) +
\hbar^{1/2}\xi$.

This is justified because when we apply a function of $x$, $f(x)$, to a squeezed state $\varphi_{k}(A, B, a, \eta, \hbar, x)$, we can
develop $f(x)$ in Taylor series, up to order $l$, around the centre of the packet, getting a remainder which, in norm, is of order
$\hbar^{l/2 +1}$ (Hagedorn 1994 and references therein).  

At this point, we make the Ansatz that
\begin{eqnarray}
\fl \tilde{\psi}(x, y, \xi; t) = \exp\Big(\rmi S(t)/\hbar\Big)\exp\Bigg[\frac{\rmi\eta(t)^{T}G_{M}(a(t))\xi}{\hbar^{1/2}}\Bigg]F(x, \hbar y)
\nonumber \\
\times g_{M}(a(t))^{-1/4}(\tilde{\psi}_{0} + \hbar^{1/2}\tilde{\psi}_{1} + \hbar\tilde{\psi}_{2} + \dots),
\end{eqnarray}
where $a(t)$ and $\eta(t)$ satisfy equation \eref{classicmultidim}, $S(t)$ is the associated action
\begin{equation}
S(t) = \int_{0}^{t} ds \frac{1}{2}\eta(s)^{T}G_{M}(a(s))\eta(s) - E(a(s)) \qquad ,
\end{equation}
and $F$ is a smooth function which has support in $x$ near the classical trajectory, and in $\hbar y$ near $0$.

Substituting this Ansatz in the equation for $\tilde{\psi}$, and keeping terms up to order $\hbar$, we can determine $\tilde{\psi}_{0}$ and
$\tilde{\psi}_{2}^{\perp}$, which, as shown in (Hagedorn 1994) are what is needed to solve the Schr\"odinger equation to lowest order in
$\hbar^{1/2}$. The calculations are lengthy and not very interesting, so we give simply the result.

The approximate solution, up to order $\hbar^{1/2}$, of \eref{schrodmultidim} is
\begin{eqnarray}
\fl \psi_{\textrm{ap}}(x, y; t) = \exp\Big(\rmi
S(t)/\hbar\Big)\exp\Bigg[\frac{\rmi\eta(t)^{T}G_{M}(a(t))\xi}{\hbar^{1/2}}\Bigg]g_{M}(a(t))^{-1/4}\hbar^{-n/4}  \nonumber \\
\times \varphi_{k}(A(t), B(t), 1, 0, 0, \xi) \Bigg\{\Phi(x, y) + \hat{r}(x)\Bigg[\rmi\eta^{T}\nabla_{x}\Phi \nonumber \\
+ \rmi\eta(t)^{T}N(a(t), y)\nabla_{y}\Phi + \eta(t)^{T}G_{M}(a(t))S(a(t), y)\eta(t)\Phi \nonumber \\
+ y^{T}\nabla_{y}V(x)\Phi + \frac{1}{a}\sum_{|p| =3}\frac{\nabla_{y}^{p}W(x)y^{p}}{p!}\Phi\Bigg]\Bigg\} \qquad ,
\end{eqnarray} 
where $\xi$ is given by \eref{xi}, $N$ and $S$ are defined in \eref{N} and \eref{S}, and $\Phi(x, y)$ is a real eigenstate of $\htrtwo(x)$,
with eigenvalue $E(x)$ of multiplicity $1$.

As before,  $\hat{r}(x)$ is the bounded inverse of the restriction of $[\htrtwo(x) - E(x)]$ to the orthogonal complement of $\Phi(x, y)$ in
$L^{2}(\field{R}{m}, dy)$.

\begin{rem} The evolution of the dispersion matrices $A(t)$ and $B(t)$ can be read from the terms of order $\hbar$ in the expansion, and
contains explicitly the metric $G_{M}$:
\begin{eqnarray}
\deriv{}{t}A(t)_{il} = \eta_{k}(t)[G_{M}\deriv{}{j}G_{M}^{-1}]_{ki}(a(t))A(t)_{jl} + \rmi [G_{M}^{-1}(a(t))B(t)]_{il} \\
\fl \deriv{}{t}B(t)_{il} = \frac{\rmi}{2}\eta(t)^{T}[G_{M}(\deriv{2}{ij}G_{M}^{-1})G_{M}](a(t))\eta(t)A(t)_{jl} +
\deriv{2}{ij}E(a(t))A(t)_{jl} \nonumber \\
- \eta_{k}(t)[G_{M}\deriv{}{i}G_{M}^{-1}]_{kj}(a(t))B(t)_{jl}
\end{eqnarray}
\end{rem}

\begin{rem} The term coming from the gauge coupling
\begin{equation*}
\rmi\eta(t)^{T}N(a(t), y)\nabla_{y}\Phi
\end{equation*}
can be written, using creation and destruction operators for the normal oscillations, as
\begin{eqnarray}
\fl \rmi\eta_{j}\beta_{j}^{(\lambda, k_{\lambda}) (\nu, h_{\nu})}y_{\lambda, k_{\lambda}}\frac{\partial}{\partial y_{\nu, h_{\nu}}}\Phi =
\frac{\rmi}{2}\eta_{j}\beta_{j}^{(\lambda, k_{\lambda}) (\nu, h_{\nu})}\Big[\frac{\omega_{\nu}}{\omega_{\lambda}}\Big]^{1/2}(a_{\lambda,
k_{\lambda}}a_{\nu, h_{\nu}} - a_{\lambda, k_{\lambda}}a^{\dagger}_{\nu, h_{\nu}} \nonumber \\
+ a^{\dagger}_{\lambda, k_{\lambda}}a_{\nu, h_{\nu}} - a^{\dagger}_{\lambda, k_{\lambda}}a^{\dagger}_{\nu, h_{\nu}})\Phi \qquad .
\end{eqnarray}

Since $\beta$ is antisymmetric in $(\lambda, k_{\lambda}), (\nu, h_{\nu})$, the above expression is orthogonal to $\Phi$, as we claimed in
the introduction to this section.
\end{rem}
 
\section{Takens chaos in quantum mechanics}

When the constraining potential is not spectrally smooth, that is, roughly speaking, when the eigenvalues or the eigenfunctions of its
Hessian are not smooth, the classical motion on the submanifold $M$ shows peculiar features.

In this section we consider the quantum analogue of an example given by Takens (Takens 1980, see also Bornemann 1998) where $W$ fails to
constrain spectrally smooth.

The Hamiltonian we study is
\begin{equation}\label{takensham}
H_{\varepsilon} = \frac{p_{x_{1}}^{2} + p_{x_{2}}^{2}}{2} + \frac{p_{y_{1}}^{2} + p_{y_{2}}^{2}}{2} + \frac{1}{2\varepsilon^{2}}<R(x)y, y>,
\end{equation}
where $q = (x, y) \in \field{R}{4}$, $<\cdot, \cdot>$ is the standard scalar product in $\field{R}{2}$ and $R(x)$ is the Rellich matrix
(Kato 1995 and references therein)
\begin{equation}
R(x) = \frac{1}{4}\Bigg[ \mathbb{I} + \left( \begin{array}{cc} x_{1} & x_{2} \\
                                                                                            x_{2} & -x_{1} \end{array} \right) \Bigg]\qquad
.
\end{equation}

The eigenvalues of $R(x)$ are
\begin{equation}\label{omega}
\omega_{\pm}(x)^{2} = \frac{1}{4}(1 \pm |x|),
\end{equation}
with corresponding eigenvectors
\begin{equation}\label{eigenvec}
v_{+}(x) = \left( \begin{array}{c} \cos(\phi/2) \\
                                                   \sin(\phi/2)    \end{array} \right) \qquad
v_{-}(x) = \left( \begin{array}{c} -\sin(\phi/2) \\
                                                   \cos(\phi/2) \end{array} \right),
\end{equation}
where $\phi = \tan^{-1}(x_{2}/x_{1})$, and the branch of the inverse tangent is chosen so that $-\pi/2\leq\phi<3\pi/2$.

The eigenvectors are discontinuous along the semiaxis $\{ x: x_{1} = 0, x_{2}\leq 0 \}$, or better, they exchange place upon crossing the
cut.

\subsection{A brief review of the classical case}

To get a confining potential which is bounded from below we restrict the configuration space to 
\begin{equation}
\Sigma := \{ (x, y): |x| < 1/2 \}.
\end{equation}

With this choice, the Hamiltonian \eref{takensham} constrains the system to the submanifold
\begin{equation}
M := \{ (x, y) \in \Sigma: y = 0\}.
\end{equation}

An (almost) complete description of the limit motions when $\varepsilon \to 0$ is given by
\begin{theo}[Takens 1980, theorem 3] Let 
\begin{equation*}
W(q) = \frac{1}{2}<R(x)y, y>,
\end{equation*}
then the solutions of the equations of motion
\begin{equation}
\eqalign{\ddot{q}_{\varepsilon}(t) = -\frac{1}{\varepsilon^{2}}\nabla W(q_{\varepsilon}(t)) \\
q_{\varepsilon}(0) = 0 \qquad \dot{q}_{\varepsilon}(0) \to v_{*}}
\end{equation}
which satisfy
\begin{equation}\label{genericity}
Qv_{*}\neq 0,
\end{equation}
where $Q: \field{R}{4} \to \field{R}{2}$ is the orthogonal projector $Q(x, y) = x$, converge uniformly to the \emph{unique} solution of
\begin{equation}
\eqalign{\ddot{x}(t) = -\nabla U_{{\rm hom}}(x(t), t) \\
x(0) = 0 \qquad \dot{x}(0) = Qv_{*},}
\end{equation}  
where
\begin{equation}\label{funnel}
U_{{\rm hom}}(x, t):= \vartheta_{+}(t)\omega_{+}(x) + \vartheta_{-}(t)\omega_{-}(x).
\end{equation}
The functions $\vartheta_{\pm}$ are constant for $t \neq 0$ and can have any discontinuity in $t = 0$, provided that $\vartheta_{+} +
\vartheta_{-}$ remains constant.
\end{theo}

\begin{rem} If $Qv_{*} = 0$, the limiting behaviour is \emph{not} known. \end{rem}

\subsection{A quantum analogue}

In the quantum case, we consider the Hamiltonian
\begin{equation}
\Hh = -\frac{\hbar^{2}}{2}(\Delta_{x} + \Delta_{y}) + \frac{1}{2a^{2}\hbar^{2}}<g(|x|)R(x)y, y>,
\end{equation}
where $g \in \cinf(\field{R}{})$, $g(z) = 1$ when $|z| <1/2$, $g(z) = 0$ when $|z| > 3/5$.

We use the same squeezing factor $a$ for both transversal directions so that the eigenvalues of  $R(x)$ keep their simple form \eref{omega}. 

The quadratic form $<g(|x|)R(x)y, y>$ is non-negative, so $\Hh$ is essentially self-adjoint on $\cinf(\field{R}{4})$.

Scaling $y$ as we did in the above sections, we get
\begin{equation}\label{crossingham}
\eqalign{\Hbo = -\frac{\hbar^{2}}{2}\Delta_{x} + \htr(x) \\
\htr(x) = -\frac{1}{2}\Delta_{y} + \frac{1}{2a^{2}}<g(|x|)R(x)y, y>.}
\end{equation}

Let us suppose from now on that $|x| < 1/2$, so that $g(|x|) = 1$ (note that, in theorem \eref{hagedorn}, it is required that $\htr(x)$ has
an eigenvalue on an open set only, so this restriction is immaterial).

To calculate the spectrum of $\htr(x)$ we exploit the fact that, for every $x$, $R(x)$ is a real symmetric matrix, and can be diagonalized
by an orthogonal transformation whose form can be derived from \eref{eigenvec}, and is given by
\begin{equation}
Z(x) =  \left( \begin{array}{cc} \cos(\phi/2) & -\sin(\phi/2) \\
                                       \sin(\phi/2) & \cos(\phi/2) \end{array} \right) .
\end{equation}
It shows the same discontinuity of $v_{\pm}$, but however is defined for all $x$.

The corresponding unitary operator 
\begin{equation}
\eqalign{\widehat{Z}(x): L^{2}(\field{R}{2}_{y}) \to L^{2}(\field{R}{2}_{y}) \\
[\widehat{Z}(x)\psi](y) = \psi(Z(x)^{-1}y)}
\end{equation}
turns $\htr(x)$ into the Hamiltonian of two uncoupled harmonic oscillators,
\begin{equation}\label{uncouposc}
\widehat{Z}(x)^{\dagger}\htr(x)\widehat{Z}(x) = -\frac{1}{2}\Delta_{y} + \frac{1}{2a^{2}}\omega_{+}(x)^{2}y_{1}^{2} +
\frac{1}{2a^{2}}\omega_{-}(x)^{2}y_{2}^{2}.
\end{equation}

The eigenvalues of $\htr(x)$ are then
\begin{equation}\label{eigenvalues}
\eqalign{E_{n_{+}, n_{-}}(x) = E_{0, 0}(x) + \frac{n_{+}}{a}\omega_{+}(x) + \frac{n_{-}}{a}\omega_{-}(x)\\
E_{0, 0}(x) = \frac{\omega_{+}(x) + \omega_{-}(x)}{2a} = \frac{1}{4a}[(1 + |x|)^{1/2} + (1 - |x|)^{1/2}].}
\end{equation}

\subsubsection{The ground state}

The eigenfunction corresponding to $E_{0, 0}(x)$ is 
\begin{equation*}
\Phi_{0, 0}(x, y) = [\widehat{Z}(x)\Psi_{0, 0}](x, y) = \Psi_{0, 0}(x, Z(x)^{-1}y),
\end{equation*}
where $\Psi_{0, 0}$ is the eigenfunction of \eref{uncouposc} with the same eigenvalue.

The result, with a suitable choice of normalization constants, is
\begin{equation}\label{groundstate}
\Phi_{0, 0}(x, y) = \Bigg[\frac{\omega_{+}(x)\omega_{-}(x)}{a^{2}\pi}\Bigg]^{1/4}\exp\Big(-\frac{1}{2a}<R(x)^{1/2}y, y>\Big).
\end{equation}

The equations \eref{eigenvalues} and \eref{groundstate} tell us that both the energy and the wave function of the ground state of $\htr(x)$
are $C^{\infty}$ functions of $x$ for $|x| < 1/2$. Therefore, theorem \eref{hagedorn} can be used also in this case, and gives us a
constrained motion in the cylinder $\{(x, y): |x|<1/2, y = 0\}$, with effective potential $E_{0, 0}(x)$.

The classical trajectory we obtained is the only one which is associated, in the funnel described by \eref{funnel}, to a smooth homogenized
potential. The semiclassical limit thus singles out a specific motion, which is linked to the initial normal oscillation.

\subsubsection{The excited states}

If we consider the excited states of $\htr(x)$, we observe crossings between different eigenvalues in $x = 0$. Unlike what happens in the
classical case, however, an incoming semiclassical wave packet splits into two components only, giving rise to a bifurcation of the
motion, and not to a funnel.

For the first two excited states, for example, we have
\begin{eqnarray}
E_{0, 1}(x) = E_{0, 0}(x) + \omega_{-}(x)/a \\
E_{1, 0}(x) = E_{0, 0}(x) + \omega_{+}(x)/a,
\end{eqnarray}
(when $|x|<1/2$ we have $\omega_{+}(x)<2\omega_{-}(x)$, so the other eigenvalues remain separated from these).

The corresponding eigenfunctions are
\begin{eqnarray}
\Phi_{0, 1}(x, y) = a^{-1/2}\Phi_{0, 0}(x, y)[2\omega_{-}(x)]^{1/2}[-\sin(\phi/2)y_{1} + \cos(\phi/2)y_{2}] \\
\Phi_{1, 0}(x, y) = a^{-1/2}\Phi_{0, 0}(x, y)[2\omega_{+}(x)]^{1/2}[\cos(\phi/2)y_{1} + \sin(\phi/2)y_{2}].
\end{eqnarray}  

Clearly, the two eigenvalues coincide when $x = 0$, and are not differentiable in such point, while the eigenfunctions are not even
continuous.

Carrying out a rotation between $\Phi_{0, 1}$ and $\Phi_{1, 0}$, we can construct a smooth basis in the two-dimensional subspace generated
by them.

It is easily seen that
\begin{equation}
\left( \begin{array}{c} \Phi_{{\rm A}}(x, y) \\
                                  \Phi_{{\rm B}}(x, y) \end{array} \right):= 
\left( \begin{array}{cc} \sin(\phi/2) & -\cos(\phi/2) \\
                                    \cos(\phi/2) & \sin(\phi/2) \end{array} \right)
\left( \begin{array}{c} \Phi_{0, 1}(x, y) \\
                                  \Phi_{1, 0}(x, y) \end{array} \right)
\end{equation}
are smooth in the origin, since $[2\omega_{\pm}(x)]^{1/2} = (1\pm |x|)^{1/4} = 1 \pm \frac{1}{4}|x| + \Or(|x|^{2})$, so
\begin{eqnarray}
\Phi_{{\rm A}}(x, y) = a^{-1/2}\Phi_{0, 0}(x, y)\Big\{ -y_{1} -\frac{1}{4}y_{1}x_{1} -\frac{1}{4}y_{2}x_{2} + \Or(|x|^{2})\Big\} \nonumber
\\
\Phi_{{\rm B}}(x, y) = a^{-1/2}\Phi_{0, 0}(x, y)\Big\{y_{2} + \frac{1}{4}y_{1}x_{2} -\frac{1}{4}y_{2}x_{1} + \Or(|x|^{2})\Big\} \nonumber .
\end{eqnarray}

Note that
\begin{eqnarray}
\fl <\Phi_{{\rm B}}(x, y), \htr(x)\Phi_{{\rm A}}(x, y)>_{L^{2}(\field{R}{2}_{y})} = a^{-1}\sin(\phi/2)\cos(\phi/2)[\omega_{-}(x) -
\omega_{+}(x)]\\ 
\lo= a^{-1}\Big[-\frac{1}{4}x_{2} + \Or(|x|^{3})\Big] \neq 0 \qquad \forall x\neq 0.
\end{eqnarray}

Therefore, in Hagedorn's classification (Hagedorn 1994), this is a crossing of type $I$. The theory developed by him allows to elaborate on
the qualitative features of the propagation we mentioned above.

If the system is initially in a semiclassical state associated to the level
$E_{0, 1}$ and passes through the region of crossing, $x=0$, with a non-zero velocity (this assumption of generic crossing was already
present in Takens' theorem, \eref{genericity}) the final state is a superposition of two components, one evolving with the potential $E_{0,
1}$ and the other with the potential $E_{1, 0}$. More precisely we have
\begin{theo}[Hagedorn 1994, theorem 6.3] There is an approximate solution $\Psi(\hbar, x, y, t)$ to the Schr\"odinger equation generated by
the Hamiltonian \eref{crossingham} that satisfies
\begin{eqnarray}
\fl \Psi(\hbar, x, y, t) = \Phi_{0, 1}(x, y)\exp\big(iS^{(0, 1); -}(t)/\hbar\big)\varphi_{k}(A^{(0, 1); -}(t), B^{(0, 1); -}(t), \hbar,
a^{(0, 1)}(t), \eta^{(0, 1)}(t), x) \nonumber \\
+ \Or(\hbar^{1/2})
\end{eqnarray}
for $t \in [-T, T_{1}]$, for any $T_{1}>0$. For $t \in [T_{1}, T]$, this solution satisfies
\begin{eqnarray}
\fl \Psi(\hbar, x, y, t) = \Phi_{0, 1}(x, y)\exp\big(iS^{(0, 1); +}(t)/\hbar\big) \nonumber \\ 
\times\sum_{m}d_{m}^{(0, 1)}\varphi_{m}(A^{(0, 1); +}(t), B^{(0, 1); +}(t), \hbar, a^{(0, 1)}(t), \eta^{(0, 1)}(t), x) \nonumber \\
+ \Phi_{1, 0}(x, y)\exp\big(S^{(1, 0), +}(t)/\hbar\big) \nonumber \\
\times \sum_{|m|<|k|} d_{m}^{(1, 0)}\varphi_{m}(A^{(1, 0); +}(t), B^{(1, 0); +}(t), \hbar, a^{(1, 0)}(t), \eta^{(1, 0)}(t), x) \nonumber \\
+ \Or(\hbar^{\alpha/2}),
\end{eqnarray}
for some $\alpha > 0$.
\end{theo}

\ack
This research has been supported by the italian MURST and by the association INTAS, sponsored by European Community.

\appendix

\section{Nondegenerate critical submanifolds}\label{nondegsub}

Let $W: \field{R}{n+m} \to \field{R}{}$ be a non-negative function, and let $M=\{q\in\field{R}{n+m}: W(q)=0\}$ be a smoothly embedded
$n-$dimensional submanifold such that
\begin{itemize}
\item $M=\{q\in\field{R}{n+m}: DW(q)=0\}$;
\item the Hessian $H$ of $W$, defined as a field of linear operators $H: M \to \mathscr{L}(R^{n+m})$ by
\begin{equation}
<H(q)u, v> = D^{2}W(q)(u, v) \qquad u, v \in \field{R}{n+m} \qquad q\in M
\end{equation}
\end{itemize}
($<\cdot, \cdot>$ is the standard scalar product in $\field{R}{n+m}$) is uniformly positive definite when restricted to $T_{q}M^{\perp}$.

Then, $M$ will be called a \emph{nondegenerate critical submanifold} of $\field{R}{n+m}$ and $W$ will be called \emph{constraining} to $M$.

\section{Spectrally smooth constraining potentials}\label{specsmooth}

Let $W$ be a potential constraining to a nondegenerate critical submanifold $M$. If the Hessian $H$ of $W$ has a smooth spectral
decomposition on $M$,
\begin{equation}
H(q)=\sum_{k=1}^{r}\omega_{k}(q)^{2}P_{k}(q), \qquad q\in M,
\end{equation}
$W$ will be called a \emph{spectrally smooth constraining potential}. Here, $\omega_{k}^{2}$ and $P_{k}$ represent the (non-zero)
eigenvalues
and eigenprojections of the Hessian.

\References
\item [ ] Abraham R and Marsden J E 1978 {\it Foundations of Mechanics} 2nd edition (Reading, Massachusetts: Benjamin/Cummings Publishing
Company)
\item[ ] Bornemann F 1998 {\it Homogenization in Time of Singularly Perturbed Mechanical Systems} (Berlin: Springer)
\item[ ] Combescure M 1992 The squeezed state approach of the semiclassical limit of the time-dependent Schr\"odinger equation {\it \JMP}
{\bf 33} 3870-80
\item[ ] da Costa R C T 1981 Quantum mechanics of a constrained particle {\it \PR} A {\bf 23} 1982-87
\item[ ] \dash 1982 Constraints in quantum mechanics {\it \PR} A {\bf 25} 2893-900
\item[ ] Duclos P and Exner P 1995 Curvature-induced bound states in quantum waveguides in two and three dimensions {\it Rev. Math. Phys.}
{\bf
7} 73-102
\item[ ] Exner P 2003 Spectral properties of Schr\"odinger operators with a strongly attractive delta interaction supported by a surface
{\it Proc. NSF Summer Research Conf. (Mt. Holyoke 2002)} (Providence, Rhode Island: American Mathematical Society) pp 25-36
\item[ ] Froese R and Herbst I 2001 Realizing holonomic constraints in classical and quantum mechanics {\it Commun. Math. Phys.} {\bf 220}
489-535
\item[ ] Hagedorn G A 1994 {\it Molecular Propagation through Electron Energy Level Crossings} (Providence, Rhode Island: American
Mathematical
Society)
\item[ ] \dash 1998 Raising and lowering operators for semiclassical wave packets {\it Ann. Phys.} {\bf 269} 77-104
\item[ ] Henneaux M and Teitelboim C 1992 {\it Quantization of Gauge Systems} (Princeton, NJ: Princeton University Press)
\item[ ] Jensen H and Koppe H 1971 Quantum mechanics with constraints {\it Ann. Phys.} {\bf 63} 586-91
\item[ ] Kaplan L, Maitra N T and Heller E J 1997 Quantizing constrained systems {\it \PR} A {\bf 56} 2592-99
\item[ ] Kato T 1995 {\it Perturbation Theory for Linear Operators} 2nd corr. ed. (Berlin: Springer) 
\item[ ] Lang S 1995 {\it Differential and Riemannian Manifolds} (New York: Springer)
\item[ ] Maraner P and Destri C 1993 Geometry-induced Yang-Mills fields in constrained quantum mechanics {\it Mod. Phys. Lett.} A {\bf 8}
861-68 
\item[ ] Mitchell K A 2001 Gauge fields and extrapotentials in constrained quantum systems {\it \PR} A 042112
\item[ ] Reed M and Simon B 1978 {\it Methods of Modern Mathematical Physics IV: Analysis of Operators} (New York: Academic Press)
\item[ ] Rubin H and Ungar P 1957 Motion under a strong constraining force {\it Commun. Pure Appl. Math.} {\bf 28} 65-87
\item[ ] Schuster P C and Jaffe R L 2003 Quantum mechanics on manifolds embedded in Euclidean space {\it Ann. Phys.} {\bf 307} 132-43
\item[ ] Sk\'ala L, \v C\'\i\v zek J, Dvo\v r\'ak J and \v Spirko V 1996 Method for calculating analytical solutions of the Schr\"odinger
equation: anharmonic oscillators and generalized Morse oscillators {\it \PR} A {\bf 53} 2009-20
\item [ ] Spivak M 1979 {\it Differential Geometry} vol. 4 2nd edition (Houston, Texas: Publish or Perish, Inc.)
\item[ ] Takens F 1980 Motion under the influence of a strong constraining force {\it Global Theory of Dynamical Systems} ed Z Nitecki and C
Robinson (Berlin: Springer) pp 425-45
\item[ ] Teufel S 2003 {\it Adiabatic Perturbation Theory in Quantum Dynamics} (Berlin: Springer)
\item[ ] Ushveridze A G 1994 {\it Quasi-Exactly Solvable Models in Quantum Mechanics} (Bristol and Philadelphia: IOP Publishing)
\endrefs

\end{document}